# Superconducting gap modulations: are they from pair density waves or pair-breaking scattering?

Jia-Xin Yin[1*], Qianghua Wang[2]

[1]Department of Physics, Southern University of Science and Technology, Shenzhen, Guangdong, China.
[2]National Laboratory of Solid State Microstructures & School of Physics, Nanjing University, Nanjing, China.
*Email: yinjx@sustech.edu.cn

**Abstract**
**In his seminal work published in Acta Phys. Sin. in 1965, Yu Lu pointed out that the superconducting gap exhibits weak modulations near the pair-breaking magnetic impurity in a superconductor. In the past ten years, a series of high-resolution scanning tunneling microscopy works reported weak superconducting gap modulations in certain superconductors and explained these phenomena as pair density waves. In line with Yu's discovery, Lee DH et al. pointed out that in many cases, the interference effect of pair-breaking scattering can also lead to superconducting gap modulations in space. We will discuss the distinction and unification of these two kinds of mechanisms, as well as their relevance to recent experimental observations.**

**Pair density wave and pair-breaking scattering: distinction and unification**
Superconductivity is a magical macroscopic quantum state, and superconducting materials exhibit zero resistance and perfect diamagnetism. The microscopic mechanism for forming this quantum state is generally believed to first require electrons near the Fermi energy of the material to form pairs, with opposite momenta for paired electrons, that is, the wave vector of paired electrons is zero [1]. Under this circumstance, the pairing order parameter does not show spatial modulation characteristics, and the superconducting gap measuring the binding energy of paired electrons exhibits a uniform distribution in space. In the early 1960s, theorists Fulde P, Ferrell RA, Larkin AI, and Ovchinnikov YN predicted that under the action of a magnetic field, the total momentum of paired electrons is nonzero, presenting a finite wave vector [2,3]. This superconducting state will exhibit spatial modulation characteristics, known as the pair density wave state or the "FFLO" state. In the "FFLO" state, the superconducting gap will exhibit oscillations in space with a period of $2\pi/Q$ (Figure 1 above), where FF is a planar wave form of pair density wave: $\Delta(\boldsymbol{r}) = \Delta_0 e^{i\boldsymbol{Q}\cdot\boldsymbol{r}}$, and LO is a standing wave form of pair density wave: $\Delta(\boldsymbol{r}) = \Delta_0 \cos(\boldsymbol{Q}\cdot\boldsymbol{r})$. In the past ten years, with the development of scanning tunneling microscopy technology and the improvement of measurement accuracy, scholars have gradually observed the weak oscillatory behavior of the superconducting gap in space in various superconducting quantum materials (the oscillation amplitude accounts for about 5% of the average value of the gap) [4-12], and identified it as the pair density wave analogous to the "FFLO" state.

Also in the early 1960s, Chinese theoretical physicists Yu Lu discovered the pair-breaking behavior of magnetic impurities in superconductors, that is, the formation of localized bound states within the superconducting gap [13]. This work is widely regarded as a classic in the field of superconductivity. Rereading the classic, we found that in the appendix of Mr. Yu Lu's paper, the periodic oscillation behavior of the gap around impurities was also derived in detail (Figure 1 below), and the magnitude of this change was pointed out to be 1/15 (Figure 2). It was indeed a negligible small amount at that time, but it was consistent with the magnitude of the gap oscillation amplitude observed in the current experiments. Recently, Lee DH *et al.* further pointed out that in many cases (including magnetic impurities, non-magnetic impurities, superconducting order parameters with phase variations, magnetic superconductors, etc.), the interference formed by the pair-breaking scattering of multiple impurities can form periodic oscillations of the pairing gap [14]. Such oscillations may not involve phase modulation and are distinct from pair density waves.

In the quantum world, opposition often implies unity. The two oscillation mechanisms described above do have profound connections. For example, the magnetic field introduced in the FFLO state explicitly breaks the time-reversal symmetry [2,3]; magnetic impurities also destroy superconductivity by explicitly breaking the time-reversal symmetry, generating localized bound states and gap oscillations [13]; when the superconducting material undergoes spontaneous time-reversal symmetry breaking, non-magnetic impurities can also destroy superconductivity and excite gap oscillations [14]. Therefore, pair density waves and pair-breaking scatterings seem to have a unified intrinsic connection with the breaking of time-reversal symmetry. Furthermore, the two may even mix within the correlated system: when the impurity pair-breaking scattering is particularly strong at a specific wave vector and shows a divergent trend, it often means the instability of the Fermi surface here, and it is possible to emerge rich density wave states at the same wave vector.

**Experimental distinction**
Most of the currently published papers interpret the existing experimental data as pair density waves. Whether this is accurate has become a serious issue in the field. We provide some thought-provoking analyses of this problem from the perspectives of scanning tunneling spectroscopies and theory. In the experiment, the scanning tunneling microscope typically performs spectroscopic imaging of the gap in a small region of the superconducting sample, and the wave vector of the gap oscillation is obtained by Fourier transforming the real-space data. Another measurement method is to use a superconducting tip and a superconducting sample to implement a Josephson tunnel junction, perform spatial imaging of the Josephson current or the Josephson zero-energy conductance peak, and then obtain the wave vector of the so-called pairing electron density oscillation by Fourier transforming the real-space data. It should be noted that the pairing electron density here does not directly correspond to the superfluid density, but is a phenomenological density of the local wave function, which actually corresponds to a complex function of the superconducting gap after rigorous derivation, approximately proportional to its square value [15]. Therefore, the signal measured by the Josephson scanning tunneling microscope is still the spatial oscillation behavior of the superconducting gap, and it cannot be exaggerated as the modulation of the local superfluid density in the usual sense. Thus, both experimental methods actually measure the superconducting gap oscillation behavior, and we cannot strictly distinguish between the pair density wave and the impurity pair-breaking scattering.

The pair-breaking impurity scattering mechanism generally requires the presence of impurities or crystal defects, while the pair density wave is the behavior of the electronic structure in the pure lattice without impurities. Therefore, if gap oscillations are measured in a clean region without impurities or defects in the experiment, it is most likely due to the pair density wave. If gap oscillations are only measured near impurities or defects in the experiment, and the oscillation disappears in the clean region, it is most likely due to pair-breaking scattering; in this case, the reciprocal space of the oscillation signal generally presents a curved or circular structure, partially reflecting the geometry of the Fermi surface. When the measured gap oscillations occur near impurities and present as spot-like signals in reciprocal space, they may come from either the pair density wave or the scattering between the spot-like regions with large density of states (hot spots) in the band structure (impurity pair-breaking scattering), or a mixture of both.

Under these most difficult circumstances, if the signal really comes from a pure pair density wave, it is possible to use the Bogoliubov Fermi surface (e.g., [16]) to confirm the existence of the pair density wave. Under finite-momentum pairing, the superconducting gap does not strictly open at all Fermi wave vectors at the Fermi energy. At certain momentum positions, the superconducting gap opens next to the Fermi energy, resulting in the formation of a residual Fermi surface known as the Bogoliubov Fermi surface. The geometric shape of the Bogoliubov Fermi surface in momentum space should have a strict space-momentum correspondence with the spatial modulation of the pairing gap. If such a correspondence is detected, then the origin of the pair density wave can be identified. It should be noted that the pair density waves in the sense of LLFO are all with respect to the pairing order parameter that can be locally defined. In many strongly correlated systems, due to the local Coulomb repulsion, pairing is more likely to occur on neighboring bonds. In this case, the relationship between the pair density wave and the spectroscopic gap or the Josephson current is more complex and requires further clarification in theory and experiments. However, this complexity does not affect the accuracy of using the Bogoliubov Fermi surface to determine the pair density wave, so the latter has an advantage. Recently, we used the above principles to make a distinction between the two types of gap oscillations in the kagome superconducting materials $KV_3Sb_5$ and $CsV_3Sb_5$ using a (Josephson) scanning tunneling microscope, and attempted to associate the oscillations from the pair density wave with the possible Bogoliubov Fermi surface [17], hoping to contribute to a deeper understanding of the differences and connections between the two mechanisms.

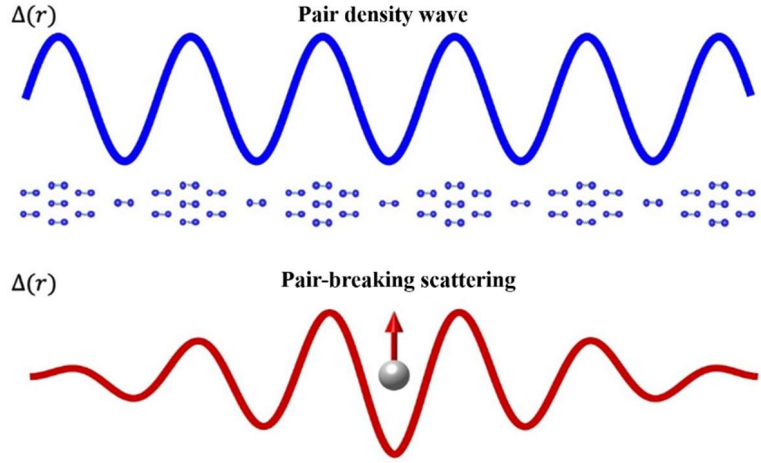

Figure 1. Superconducting gap modulations caused by pair density waves (upper panel) and pair-breaking impurity scattering (lower panel).

当 $r \gg \xi_0$ 时,可将积分限延至 $(0, \infty)$,则

$$\frac{\delta\Delta(\mathbf{r})}{\Delta_0} = \frac{b2m}{r^2}(V' \cos 2k_F r - \sin 2k_F r)\frac{\pi}{2}\sqrt{\frac{\xi_0}{r}}e^{-\frac{2r}{\pi\xi_0}}. \quad (\text{I}.9)$$

当 $r \ll \xi_0$ 时, $\cos 2kr$ 等可提出积分限,有

$$\frac{\delta\Delta(\mathbf{r})}{\Delta_0} = \frac{b2m}{r^2} \ln \frac{2\omega}{\Delta_0}(V' \cos 2k_F r - \sin 2k_F r). \quad (\text{I}.10)$$

估计在一个原胞边上能隙的变化

$$\left|\frac{\delta\Delta(r_0)}{\Delta_0}\right| \approx \frac{1}{3}\frac{V}{\varepsilon_F}\frac{1}{1+V'^2}r_0 k_F.$$

其中 $r_0 k_F = \sqrt{\frac{9\pi z}{4}}$, $V' = \frac{V}{\varepsilon_F}\frac{3\pi z}{4}$, $z$ 是价电子数. 若設 $\frac{V}{\varepsilon_F} = \frac{1}{5}$, $z = 3$, 则 $\left|\frac{\delta\Delta(r_0)}{\Delta_0}\right| \approx \frac{1}{15}$, 这确是一个小量.

Figure 2. Results on superconducting gap modulations near the magnetic impurities by Yu L from Ref. [13].